\documentclass[10pt]{article}
\usepackage{palatino}
\usepackage{amsmath}
\usepackage[a4paper, total={6.6in, 10in}]{geometry}
\usepackage{graphicx}
\usepackage[breaklinks=false,colorlinks=true,linkcolor=black,urlcolor=black,citecolor=black]{hyperref}
\usepackage{abstract}

\usepackage{multicol}
\usepackage{float}
\usepackage{braket}
\usepackage{csquotes}

\usepackage{titlesec}

\titleformat*{\section}{\large\bfseries}
\titleformat*{\subsection}{\normalsize\bfseries}

\usepackage[font=footnotesize]{caption}

\usepackage{subfig}
\usepackage{cite}
\begin{document}
		\title{5G Network Slicing with QKD and Quantum-Safe Security }
	\date{}
	
	\author{Paul Wright$^1$, Catherine White$^1$, Ryan C. Parker$^{1 *}$, Jean-S\'ebastien Pegon$^2$, \\ Marco Menchetti$^1$, Joseph Pearse$^3$, Arash Bahrami$^3$, Anastasia Moroz$^1$,\\ Adrian Wonfor$^4$, Timothy P. Spiller$^3$, Richard V. Penty$^4$, Andrew Lord$^1$}
	\maketitle
	
	\vspace{-1cm}
	\begin{center}
		$^{1}$\textit{BT Labs, Adastral Park, Ipswich, U.K.}
		
		$^{2}$\textit{ID Quantique SA, Geneva, Switzerland}

		$^{3}$\textit{York Centre for Quantum Technologies, Department of Physics, University of York, York, U.K.}
		
		$^{4}$\textit{Department of Engineering, University of Cambridge, Cambridge, U.K.}
		
		$^{*}$\textit{Corresponding author: Ryanparker1447@gmail.com}
	\end{center}
	
	\begin{abstract}
		
		We demonstrate how the 5G network slicing model can be enhanced to address data security requirements. In this work we demonstrate two different slice configurations, with different encryption requirements, representing two diverse use-cases for 5G networking – namely, an enterprise application hosted at a metro network site, and a content delivery network. 
		
We create a modified software-defined networking (SDN) orchestrator which calculates and provisions network slices according to the requirements, including encryption backed by quantum key distribution (QKD), or other methods. Slices are automatically provisioned by SDN orchestration of network resources, allowing selection of encrypted links as appropriate, including those which use encryption with standard Diffie-Hellman key exchange, QKD or quantum-resistant algorithms (QRAs), as well as no encryption at all. We show that the set-up and tear-down times of the network slices takes of the order of 1-2 minutes, which is at least an order of magnitude improvement over manually provisioning a link today.  
\\[0.08in]
		
	\end{abstract}
\begin{multicols}{2}
		\section{Introduction}\label{Sec:Introduction}
		The recent introduction of 5G networks for commercial use promises to deliver increased bandwidth to customers, enabling faster speed connections, as well as lower-latency communications, the ability to meet Quality of Service demands, and many other service improvements. This opens up the possibility for far greater connectivity of devices than ever before. 
		
The benefits brought by 5G are as a result of the converged architecture, which is the core of 5G networks; resources are placed as close to the edge of the network as possible (i.e. as far away from the core network as can be), thus offering lower-latency services via so-called edge-computing \cite{Ref1}. Making use of the placement of the resources at the edge of the network, and the fact that these resources are used more efficiently (with some sharing of compute resource, for example), are use-cases such as content delivery networks (CDNs) and edge-compute, automated vehicles and remote operations, as well as the monitoring and control of large-scale Internet of Things (IoT) networks, such as smart meters and distributed power generation. 
		
Due to the fact that there are a wide variety of new use-cases which are enabled by 5G technology, the network has had to be designed such that it can cope with a wide range of heterogeneous requirements, such as latency, reliability, security, and more. Consequently, network slicing\cite{Ref2} is utilised, and plays a key role within making 5G networks suitably flexible \cite{Ref3}. 
		
By effectively multiplexing separate virtualised networks over common physical infrastructure, network slices are made, and can be provisioned with different resources. For example, a network slice providing communications for an automated vehicle will require very low latency, but a fairly low bandwidth, compared to high-definition video streaming which is more reliant on large bandwidth and less on latency \cite{Ref4}. Both of these use-cases can be delivered on the same physical infrastructure by separating these into separate virtualised networks through network slicing. 

    \begin{figure*}[t]
		\centering 
		\scalebox{0.75}
		{\includegraphics{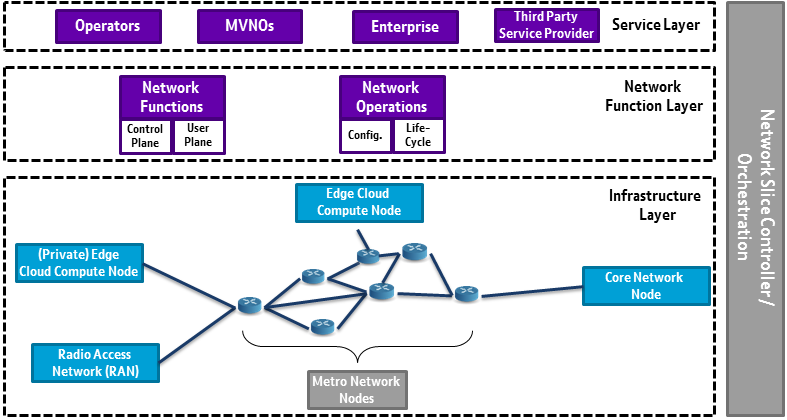}}
		\caption{A generic schematic to illustrate network slicing, orchestrated by a network slice controller within an exemplar 5G network \cite{Ref2,Ref18}.}
		\label{Fig:Generic_5G_Network_Slicing_Diagram}
	\end{figure*}
		
Network slicing is reliant on software-defined networking (SDN) and network function virtualisation (NFV). The use of SDN in transport networks is already a well-researched topic \cite{Ref19} and extending the concept for use in 5G networks has already been considered  \cite{Ref20}. NFV allows network slices to be made via virtual machines (VMs), which are then connected together across the network via SDN orchestration \cite{Ref5}; SDN is used to flexibly configure network slices, as well as reserving resources for the wide range of use-cases possible via orchestration carried about by a network slice controller (as illustrated in Fig. \ref{Fig:Generic_5G_Network_Slicing_Diagram}).  This SDN orchestration is vital within this work, as it is used to dynamically control the type of encryption deployed for each network slice. 

5G networks often provide some encryption of traffic, for example between the user equipment and the eNodeB, or to the Secure Gateway (SeGW). In general, however, they may not provide intrinsic encryption of data traffic between the end user and cloud applications, instead relying on over-the-top encrypted sessions in the application layer. This often places a responsibility on the end user and/or application developer to maintain security. \cite{Ref5}.  Application layer encryption of traffic will always be an important method to obtain end-to-end security, but 5G networks involve critical links within the tiered resources over which large concentrations of secure application traffic may flow, such as between the aggregation and metro nodes. These critical links could be attractive targets for eavesdroppers; and so we suggest that network operators consider providing intrinsic Layer 1 (L1) encryption for them. Providing L1 encryption, as in this paper, also reduces the opportunity for an eavesdropper to perform traffic analysis using packet headers.
		
A vital prerequisite for strong encryption is secure key exchange. Today’s standard key exchange algorithms (such as Diffie-Hellman and RSA) are thought to be vulnerable to attacks by large-scale quantum computers. As such, there are two possible routes for avoiding this future threat: quantum-resistant algorithms (QRAs), such as those being developed under the NIST program \cite{Ref6}, and quantum key distribution (QKD). 
		
	\begin{figure*}[t]
		\centering 
		{\includegraphics[width=0.6\linewidth]{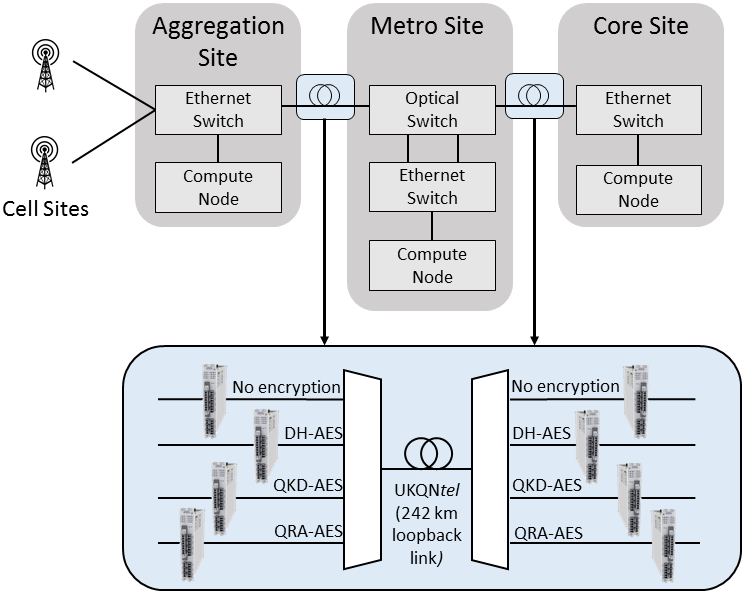}}
		\caption{The network test-bed configuration for the implementation of 5G network slicing, with varying levels of security provided.}
		\label{Fig:5D_Demo_Node_Testbed}
	\end{figure*}
		
Whereas QRAs for key exchange would be reliant on strong mathematical testing and/or proofs to safeguard against the increased compute power of a large-scale quantum computer, QKD is based upon the fundamental laws of quantum physics, and if implemented properly is secure against any future computational threat. QKD utilises quantum states encoded on photons to agree a key between users with information theoretic security (ITS). ITS implies that we are able to calculate the statistical likelihood that an eavesdropper holds any information on the key, and show that this has been reduced to an infinitesimally small probability. We emphasise that QKD is secure against any future computational threat, be that classical or quantum, whereas QRAs may be insecure against a future quantum hacking algorithm, which is yet to be discovered.
		
QKD requires an initial authentication step, which may be based on symmetric authentication where pre-shared key exists, but if this is not the case then other approaches, such as authentication algorithms based on QRAs, may be needed for this first-time authentication. If an asymmetric algorithm is used for the initial authentication step - even standard PKI - then the time-window of vulnerability of this authentication is limited, as once a key has been established using QKD it does not then matter if the initial authentication is subsequently broken, because the QKD key material has no algorithmic link to the material that was used to authenticate the QKD exchange. 
		
To protect data for which there is a need for privacy or intellectual property retention over a time-scale of years, we anticipate that network application designers will select a cryptographic suite based on QRAs. However, for valuable and/or sensitive data, further long-term key security can be provided by QKD for key exchange, in combination with QRAs for encryption and authentication. Sections of a single network slice may have different security requirements, for example where data is time sensitive and cached within the network, such as in CDNs, or where data from multiple devices is aggregated; the level of security is another parameter of the connection which it would be useful to be able to control as part of a network slice. In this way, we create the capability to dynamically control the type of encryption used for separate data channels in 5G networks. 
		
Using network slicing to control encryption is relatively novel, but nevertheless has already been considered theoretically in \cite{Ref7} and \cite{Ref8} by utilising QKD in tandem with a QRA (specifically, a QRA version of Elliptic-Curve Cryptography), and has also performed experimentally over the Bristol City 5G UK Test Network in the works of \cite{Ref9,Ref10,Ref11}, by applying QKD to 5G networking. Moreover, in \cite{Ref12}, proof-of-transit of the 5G data traffic is demonstrated, using cryptographic techniques with QKD over the Madrid Quantum Network \cite{Ref13} – this network has also been used to demonstrate securing the management of the SDN control plane through QKD in \cite{Ref14,Ref15}.  
However, what differentiates our work is that we dynamically control the \textit{type} of encryption – Diffie-Hellman+AES, QRA+AES, QKD+AES, or no encryption at all – to address the realistic scenario in which different data packets in a 5G network will have varying security requirements. We note here that the symmetric encryption algorithm used in this work is the Advanced Encryption Standard (AES) with 256 bit keys, from QKD, Diffie-Hellman or a QRA for key exchange.  AES-256 is currently thought to be "quantum-safe"\cite{Ref6}, in that even a large-scale quantum computer will be unable to crack this method of encryption with an exponential speed-up, unlike Diffie-Hellman or RSA asymmetric algorithms which are susceptible to this type of cryptanalysis. 
		
Within this work we experimentally demonstrate 5G network slicing to dynamically control the type of encryption (and therefore the level of data security) over existing commercial telecommunications infrastructure, to represent the possibility of supporting the variety of potential new use-cases born through 5G networks, which will inevitably have diverse security requirements. More specifically, we test an experimental proof-of-concept of two potential use-cases – an enterprise application hosted at a metro site in the network, and a CDN use-case.
		
This paper is organised as follows: in Section \ref{Sec:Methodology} we describe our 5G network topology and design, and methodology behind our proof of concept demonstration, before discussing the results in Section \ref{Sec:ResultsDiscussion}. Section \ref{Sec:ResultsDiscussion} is divided into subsections in which we first address the two network slice configurations separately (Subsections \ref{Subsec:UseCase1} and \ref{Subsec:Usecase2}), before moving to present results regarding the timing (namely the provision and deprovision times) of each network slice in Subsection \ref{Subsec:Timing}.

		\section{Network Configuration}\label{Sec:Methodology}
		
		Within this section we describe the network set-up behind the test-bed configuration of our 5G network slicing prototype, with dynamically-controlled encryption. 

Fig. \ref{Fig:5D_Demo_Node_Testbed} schematically describes the architecture of the representative network test-bed used within this work. There are four node types in this network – cell, aggregation, metro, and core. Traffic flows from the cell sites to the core site, via use of Ethernet switches and optical switches to switch data channels between links using different forms of encryption (note that the links required carry the quantum channel to establish QKD were static, which avoided lengthy restartup times that are often incurred with QKD systems). In this work we use the UKQN\textit{tel} infrastructure, which is a section of the UK Quantum Network, containing intermediate trusted nodes for QKD link handover and classical amplification (for further detail, see \cite{Ref16}), as this has QKD-capable networking over a 121 km link from BT Research Labs in Ipswich (Adastral Park) to Cambridge. As described in \cite{Ref16}, this link consists of four QKD sections, linked by 3 trusted nodes held in BT exchanges, and an end-to-end key is established by passing a key (here generated by a QRNG) over each QKD section using one-time-pad encryption, where it is relayed to the far endpoint via the trusted nodes.  Available for interconnections over this infrastructure are 5$\times$100G channels on a coherent dense wave-division multiplexing (DWDM) system looped back over the 121 km optical fibre link (242 km in total – see Fig. \ref{Fig:5D_Demo_Node_Testbed}).	

		\begin{figure*}[t]
		\centering 
		{\includegraphics[width=\linewidth]{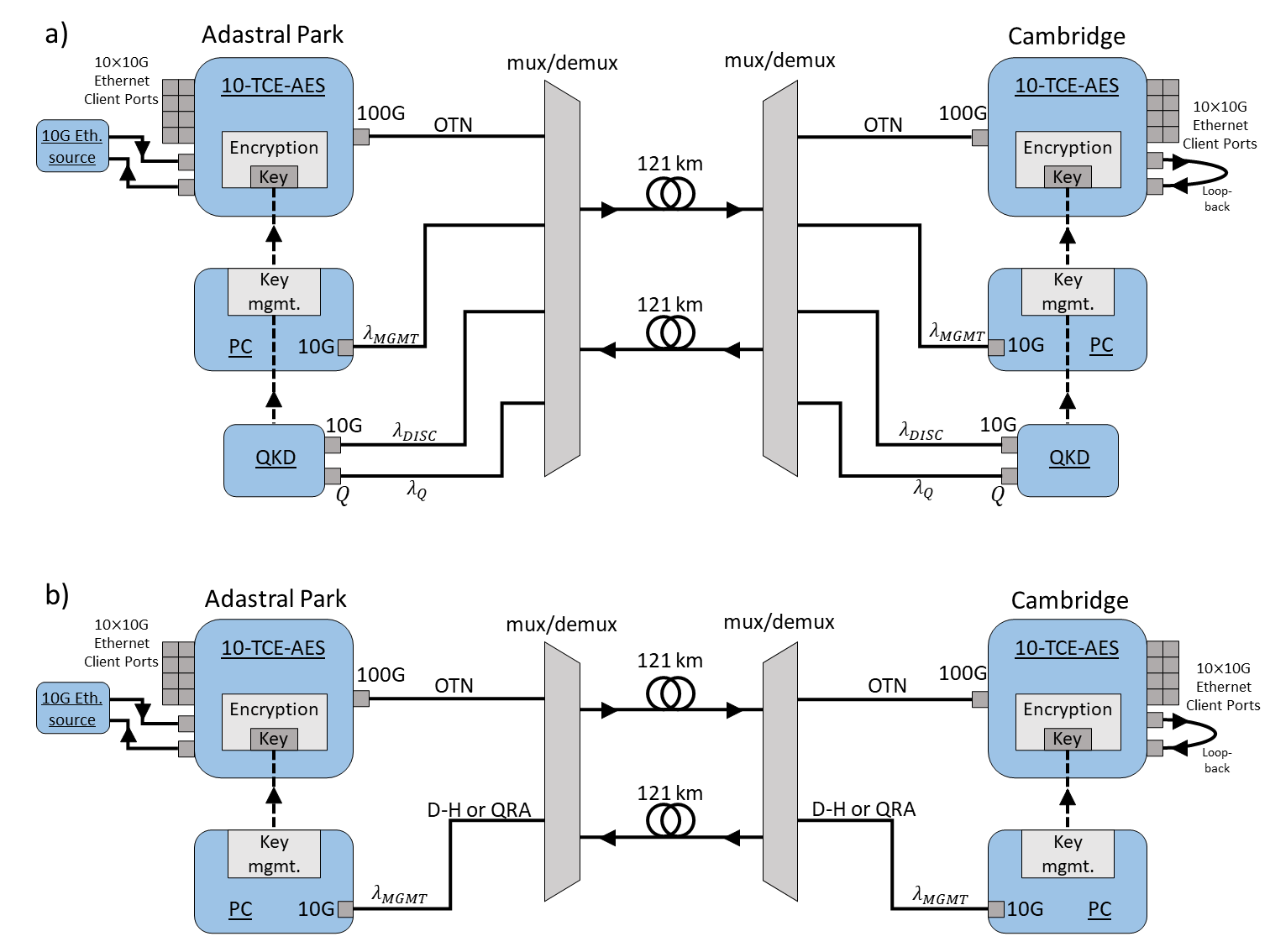}}
		\caption{Exemplary network diagrams to show the connectivity from the Adastral Park and Cambridge network nodes, illustrating the use of various encryption methods: a) QKD+AES encryption, b) DH+AES or QRA+AES encryption. The wavelengths of the various channels are denoted as follows: $\lambda_{\text{\textit{MGMT}}}$ = management wavelength, $\lambda_{\text{\textit{DISC}}}$ = QKD discussion channel wavelength, $\lambda_{\text{\textit{Q}}}$ = quantum key transmission wavelength. OTN = optical transport network, mux/demux = wavelength division multiplexing (WDM) multiplexer/demultiplexer.}
		\label{Fig:Adastral_Cambridge_Circuit_Connectivity}
	\end{figure*}		

Each of the five 100G channels within this link provides 10$\times$10G client Ethernet ports, and all interconnections between 5G network sites are 10G. There is no segregation of encryption between 10G clients on the same 100G channel (one encryption key per 100G channel, refreshed at 3s intervals). This refresh rate of 3s is not a requirement, however we are refreshing as fast as reliably possible to demonstrate optimum performance, and this corresponds to 300 Gb of data per key. In other implementations it might be preferred to have a separate encryption key per client port, but this would not affect the principle of the Network Slicing Orchestrator approach demonstrated here. The network testbed could be extended to provide links secured with keys derived from a combination of exchange methods (for example a key derived from a cryptographic hash function on 2 or more constituent keys from different methods such as QKD and QRA). However, the main purpose of this work was a proof of concept of the dynamic provision of L1 connections from links with different physical and security characteristics, and so we did not provision links to use combinations of keys at this stage, although we note that the latest version of the 10-TCE-AES card can internally derive a key from a combination of DH and a QKD key received from an external QKD system.
		
	\begin{figure*}[t]
		\centering 
		{\includegraphics[width=0.3\linewidth]{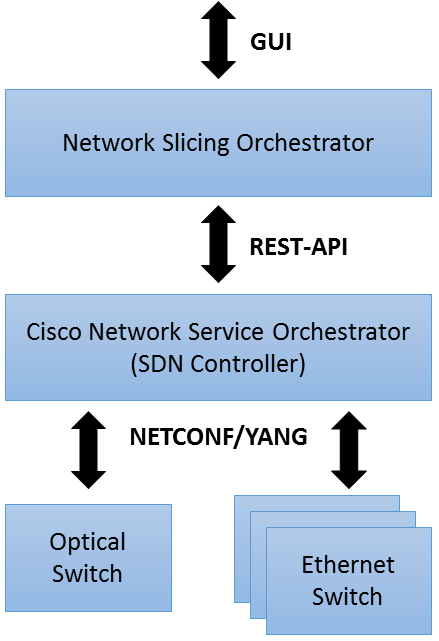}}
		\caption{Layers of network control within this experiment, showing the orchestrator, controller and device layers and the interfaces between them.}
		\label{Fig:Workflow}
    \end{figure*}

	\begin{figure*}[t]
		\centering 
		{\includegraphics[width=\linewidth]{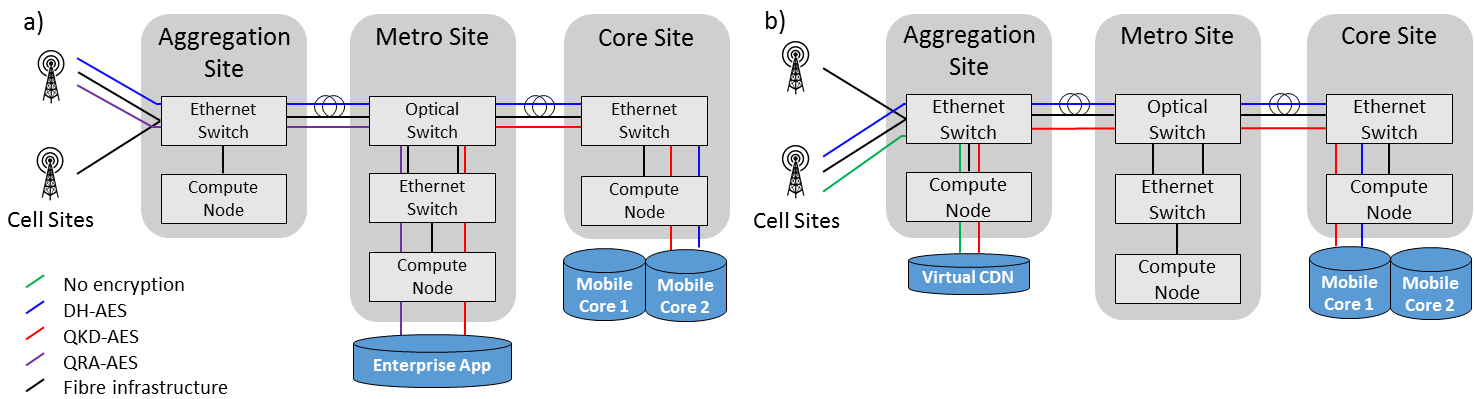}}
		\caption{Network configurations of a) Use-Case 1, representing an enterprise app hosted at a network metro site, and b) Use-Case 2, representing a content delivery network (CDN), hosted at the network aggregation site.}
		\label{Fig:5G_Testbed_UseCases_2}
    \end{figure*}
    
 Three channels are configured to provide: no encryption; standard Diffie-Hellman key exchange with Advanced Encryption Standard (DH+AES); a prototype QRA for key exchange, specifically an NTRU implementation provided by the OpenQuantumSafe library \cite{Ref17} with AES (QRA+AES). The remaining two 100G channels are in the default configuration for the UKQN\textit{tel} link (256 bit AES, with keys provided via QKD, referred to herein as QKD+AES, using the Clavis 3 ID Quantique QKD devices). Two exemplary network circuit schematics are shown in Fig. \ref{Fig:Adastral_Cambridge_Circuit_Connectivity} to illustrate the specific connectivity between Adastral Park and Cambridge with the various encryption schemes utilised in this work. In this case, a modified test firmware for the card was supplied by ADVA together with a primary/secondary SSH script for retrieving and refreshing AES keys used by the 10-TCE cards with from a third party system, which was a proof-of-concept key management system running as a TCP service and implemented on Centos 6.2. The key management system holds caches of each type of key (e.g. QKD, QRA), managed by key ID. Each client port pair on the 10-TCE-AES cards was configured statically to use a particular key exchange method, and the SDN algorithm therefore had a limited set of encrypted links using each key type to select between, and did not dynamically reconfigure the type of key used on each link. 
		
The ADVA 10-TCE encryption cards that were used for data transmission have two available models: one which supports encryption (10-TCE-AES, see Fig. \ref{Fig:Adastral_Cambridge_Circuit_Connectivity}), and one which does not (10-TCE). The resource limitations on encrypted links are therefore dependent on the hardware available. Similarly, adding QKD to an encrypted link is limited by available installed hardware, however, it may be possible to route traffic which does not strictly require encryption over free encrypted links. The delay introduced by the 10-TCE-AES is ${15\, \mu\text{s}}$ (${4\, \mu\text{s}}$ in the card, and ${11\, \mu\text{s}}$ in the CFP module which applies Forward Error Correction) - this figure is the same for both the 10-TCE and 10-TCE-AES (encrypted) card.
		
To demonstrate the ability of our orchestrator to create very diverse network slice requirements we added a further illustrative variation, namely between DH+AES and QRA+AES. However, in practice a network operator would likely select a network policy which always applies one, or both, of these techniques in addition to available QKD hardware. We view the QKD+AES encrypted links as offering the highest level of security, and note that in some implementations, since the main extra cost is for the QKD hardware, these may be implemented as QKD plus another method of key exchange in a single link. 

Central to this experiment is the use of SDN control and orchestration technologies. All of the network devices utilised within this demo have a YANG device model, and their configuration can be changed by issuing requests via a NETCONF interface. The standard YANG model of the optical and Ethernet switches have been used without any extensions. Network devices are registered with a single instance of the Cisco Network Services Orchestrator which acts as the SDN Controller, and the orchestrator communicates with the SDN controller via a REST-API. The hierarchy of the different control layers used in this experiment is illustrated in Fig. \ref{Fig:Workflow}. Each slice is broken down into three connections: cell site to core site (for control plane traffic), cell site to compute site, and compute site to core site. To achieve the required network flexibility, Layer 2 (L2) switches are used at each site. The optical switch at the metro site provides necessary flexibility for allocation of the links with different security levels to different tasks. 

	\begin{figure*}[t]
		\centering 
		{\includegraphics[width=\linewidth]{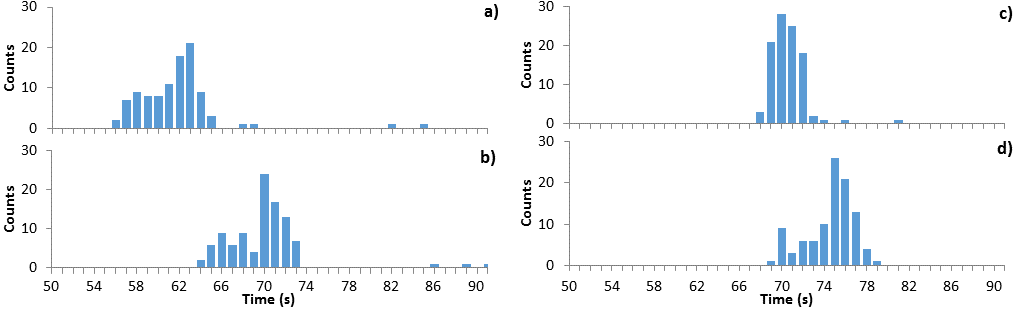}}
		\caption{Histograms showing slice set-up times over 100 runs. a) Use-Case 1 provision times, and b) deprovision times, c) Use-Case 2 provision times, and d) deprovision times.}
		\label{Fig:Histogram_Plots}
	\end{figure*}
	
This approach allows a network operator to specify the properties of the new network slice required, through a portal or application programmable interface (API). The entity providing this is a custom Network Slicing Orchestrator, which has been modified to include security requirements in both the network slice descriptor and what the underlying network can provide. The Network Slicing Orchestrator has the full end-to-end view of the network and understands the requirements for network slices, as well as performing the routing and resource allocation. 
		
For each connection in a slice, the required security level (non-encrypted, DH+AES, QKD+AES or QRA+AES is specified, along with more traditional slice parameters such as bandwidth, latency and compute requirements. Once the properties for a slice are submitted, the Network Slicing Orchestrator determines a suitable route through the network and checks whether sufficient network and compute resource are available, whilst also ensuring that the links selected meet the security requirements specified in the initial slice request. The orchestrator achieves this by allocating a security metric to each link which then is used as part of the path computation element. The network operator can then submit their request for the slice to be activated and the orchestrator then issues the configuration commands to the network devices.

		\section{Results \& Discussion}\label{Sec:ResultsDiscussion}
		We trialled two use-cases for 5G network slicing encryption. Two slice configurations are shown in Fig. \ref{Fig:5G_Testbed_UseCases_2}, based on use-cases, and in the following subsections we discuss the network topology of each use-case separately before moving to present further results.  

\subsection{Use-Case 1: Enterprise App}\label{Subsec:UseCase1}
Use-Case 1 is an enterprise app hosted at the metro site. The enterprise app processes data coming from user equipment (UE) which is connected to the cell sites. 
		
The link from each cell site to the metro site is secured with post-quantum security via use of a QRA, a solution which scales well. Premium QKD+AES encryption is selected for the link which passes aggregated data from the metro site to the mobile core node; this could be a prime target for a malicious eavesdropper, and therefore would benefit the most from the highest level of data security. Standard software-based key-exchange algorithms (Diffie-Hellman implemented within TLS 1.2) are chosen as sufficient to protect the control plane, operating from the cell site to the core site, which is considered to require only short-lived security of encryption.

\subsection{Use-Case 2: CDN}\label{Subsec:Usecase2}
Use-Case 2 is a CDN, in which the delivery sites are placed close to the network edge, at aggregation sites, in order to reduce the load within the core of the network. The scenario is that sensitive data (such as pre-released video content or software packages) is delivered securely to the CDN, and an eavesdropper would place high value in retrieving this data ahead of the official release. 
		
The delivery of the content to the CDN is via an encrypted link based on QKD, while no encryption is provisioned between the aggregation node and the cell site, since after the data has been released it no longer needs to be protected. Again, we deploy standard DH+AES key exchange and encryption to the control plane traffic, from the cell site to the core site, as we did for Use-Case 1.
		
\subsection{Timing}\label{Subsec:Timing}
		Fig. \ref{Fig:Histogram_Plots} shows histograms to quantify the time taken to set-up (provision) and tear-down (deprovision) the network slices, in both use-cases. 
		
Fig. \ref{Fig:Histogram_Plots} shows that the distribution of times to set-up and tear down each of the two slices is, in each use-case, between 1 and 2 minutes. Comparing this to the time it would take a skilled network engineer to design a service and manually configure each individual device or system of an already existing network which would typically be measured in hours, this is a significant improvement that is a benefit to telecommunications operators. In Use-Case 2 the slice takes longer to provision/deprovision as it has an additional network element to provision (namely, a metro node Ethernet switch), which is not needed in Use-Case 1. 
		
\begin{figure*}[t]
		\centering 
		{\includegraphics[width=\linewidth]{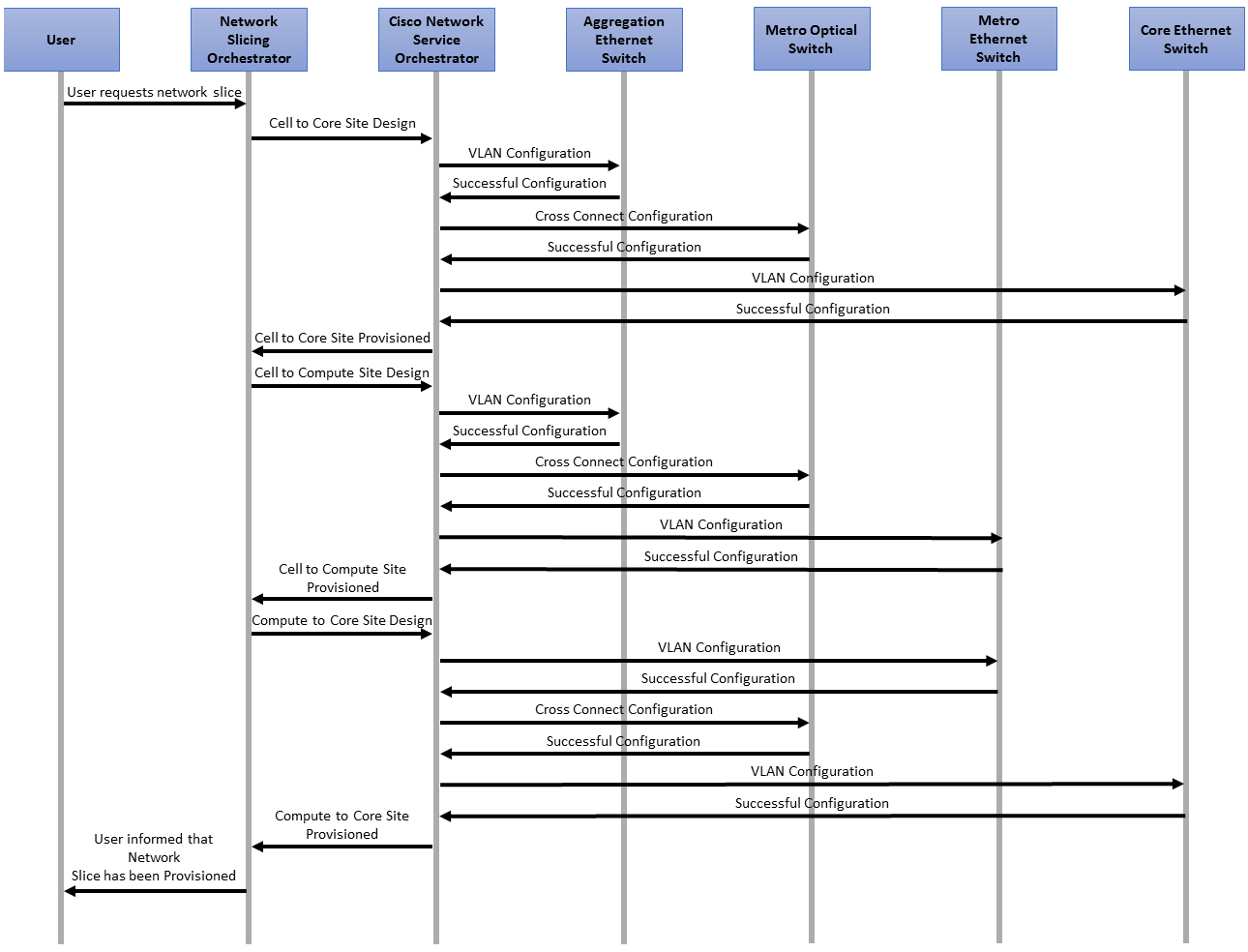}}
		\caption{Swim lane diagram showing the interactions between the different entities for a provisioning operation.}
		\label{Fig:Swimlane_Flowchart}
	\end{figure*}

Each network configuration step is made in sequence (see swim lane diagram for a provisioning operation, Fig. \ref{Fig:Swimlane_Flowchart}), allowing for efficient roll-back if there is a problem. This sequential build-up of the slice increases the time taken to set it up (there is no parallel allocation or configuration of resources), but since the network configuration is locked by the orchestrator which only allows one change at a time, this approach would reduce race conditions and conflicts if this system were to be extended to support multiple simultaneous slice requests. 
		
\section{Conclusion}\label{Sec:Conclusion}
As highlighted throughout this work, there are use-cases within network slicing and 5G networks that would greatly benefit from flexible selection of network encryption.  Two such use-cases we demonstrate in this work are metro-site-hosted enterprise apps and content delivery networks, however there are many potential applications such as CAVs (connected and automated vehicles) communications, smart factories, connecting distributed research facilities with high-value intellectual property, and more. Moreover, the dynamic nature of this work also lends itself to applications with time variable demand, such as setting-up highly secure links for daily, or more frequent, back-up of data. 

In this implementation, the provisioned links have static security parameters (i.e. type of encryption and key-exchange), and selection is made from the available static links. There is a parallel with the static nature of other parameters of links, such as latency and bandwidth. However, this design could be further developed such that those links that support encryption could be dynamically configured to select algorithms and key-exchange methods.  Such ‘cryptoagility’ could improve provisioning and response times, e.g. applying critical updates to address newly discovered vulnerabilities, or provisioning a secure link to meet particular security requirements. However, the design principles involved with rapid dynamic changes to cryptographic protocols on a network link are a topic for further research as there is a risk that the flexibility could also introduce vulnerabilities.
		
For future-proof security, the secure link options will need to include quantum-safe methods such as the quantum-resistant algorithms selected by the NIST post-quantum cryptography program, and QKD as demonstrated here, such that the customer, or network operator, is able to select the encryption level accordingly, based on the type of traffic. The security requirements of a 5G application can be included in the resource selection criteria of a 5G Network Slicing Orchestrator. This approach could help operators make maximum utilisation of premium security resources such as high speed, encrypted links and QKD.

\section*{Funding}The UK Engineering and Physical Sciences Research Council (EPSRC) (EP/N015207/1, EP/M013472/1, EP/N509802/1).

\section*{Acknowledgments}The authors thank the UK Quantum Communications Hub and ADVA for invaluable support.

\end{multicols}

\end{document}